\documentclass[aps,pre,reprint,superscriptaddress,amsmath,amssymb]{revtex4-2}

\usepackage{graphicx}
\usepackage{dcolumn}
\usepackage{bm}
\usepackage{hyperref}
\usepackage{color}
\usepackage{booktabs}

\begin{document}

\title{Human population dynamics as a Bayesian inverse transport problem}

\author{Chong Qi}
\email{chongq@kth.se}
\affiliation{Department of Physics, KTH Royal Institute of Technology, SE-10691 Stockholm, Sweden}

\date{\today}

\begin{abstract}
Many open problems across physical, biological, and engineered systems involve non-equilibrium transport processes where the governing conservation laws are known, but the underlying constitutive relations remain latent and time-varying. Conventional data-driven approaches like deep neural networks capture statistical patterns but routinely violate fundamental mass conservation. Here, we introduce a unified Bayesian inverse transport framework that resolves this by embedding Bayesian Neural Networks (BNNs) directly within exact partial differential equations in age-time space. By evaluating this framework on complex, real-world human cohort advection across China, Japan, and South Korea, we demonstrate that this physical constraint enables consistent uncertainty propagation and missing-data reconstruction from sparse observations. Beyond demography, this framework provides a generalizable foundation for observing and forecasting non-equilibrium boundary dynamics across various fields.
\end{abstract}

\maketitle

\section{Introduction}

Many dynamical systems in physics are naturally described by transport equations. Examples include neutron transport in nuclear reactors, radiative transfer in stellar atmospheres, charge carrier transport in semiconductors, and stress-strain relations in continuum mechanics. Once the transport mechanism is established, the scientific problem shifts from determining the equation of motion to identifying the system-specific constitutive relations, such as cross sections, opacities, or thermal conductivities. In other words, while the governing transport equation is assumed to be known, the constitutive relations must be inferred from observations.

In this context, human demographics represent a highly non-equilibrium advective flow governed by latent boundary conditions. Under this transport-theoretic view, individuals move deterministically through age space with unit velocity, while births introduce new individuals at the boundary and mortality acts as a time-varying internal sink. The system is governed by a first-order partial differential equation (PDE) for age-structured population transport~\cite{mckendrick1926, vonfoerster1959}. Here, age $a$ serves as the transport coordinate, analogous to spatial position in a one-dimensional advection equation, with new births acting as a boundary influx at $a=0$. In practice, these continuous dynamics are often approximated using a discrete matrix formulation~\cite{leslie1945} that maps age-specific survival and fertility into a linear state transition.
Existing demographic models generally treat fertility and mortality as prescribed inputs. The cohort-component method projects an observed age distribution forward using deterministic, scenario-based schedules, lacking a native representation of uncertainty~\cite{preston2001}. Ref.~\cite{leecarter1992} factorizes age-specific vital rates via low-rank decompositions with univariate time-series models. The United Nations Population Division applies hierarchical Bayesian models to scalar summary indicators including the total fertility rate (TFR) and life expectancy~\cite{alkema2011, raftery2012, raftery2014}. Machine-learning approaches often bypass mechanisms altogether and learn population trajectories directly from data. These approaches share common limitations: (i)~they treat age-specific vital rates as independent parameters rather than a coupled two-dimensional age-time field, losing cross-cohort correlation structure; (ii)~they do not enforce demographic consistency as a hard physical constraint, allowing mass conservation to be violated; and (iii)~they rely on empirical projections that lack physical interpretability and mechanistic feedback.

In this work, instead of prescribing explicit functional forms for fertility and mortality, we parameterize both using Bayesian neural networks and, within the Bayesian inverse problem framework~\cite{stuart2010}, embed them inside the exact transport PDE so that every posterior sample automatically satisfies demographic mass conservation. This formulation unifies historical reconstruction, forecasting, and uncertainty quantification in a single framework: missing data are recovered through Bayesian inference rather than interpolation, and future projections follow from propagating posterior samples through the transport dynamics. We sample the posterior over neural network weights using the No-U-Turn Sampler (NUTS)~\cite{neal2011, hoffmangelman2014}, a Hamiltonian Monte Carlo method previously applied by the present authors to Bayesian inference of atomic masses~\cite{storbacka2026}. While several recent physics-inspired models have used nonlinear dynamics to predict global population trajectories~\cite{sojecka2024, yakovenko2025, zaccone2026}, our focus is on the detailed, age-structured density distribution $\rho(a,t)$, which is the critical state variable governing future labor capacity, dependency dynamics, and structural stability. We apply this approach to three East Asian economies---China, Japan, and South Korea---which currently have the lowest fertility rates among major economies and represent extreme cases of demographic contraction and labor support stress.

\section{Methodology}

The foundation of demographic modeling rests on exact conservation laws. These dynamics were originally formulated as a continuous age-structured transport PDE~\cite{mckendrick1926}, which was later generalized across cellular and ecological contexts~\cite{vonfoerster1959}. In the statistical physics of society and evolutionary biology, demographic fluctuations and population dynamics have also been modeled using macroscopic advection-diffusion equations~\cite{hernandez2004, olla2012} as well as agent-based or Monte Carlo simulations, most notably the Penna bit-string model of aging and mutation accumulation~\cite{penna1995}---an approach that underscores the long tradition of applying physical and mathematical methods~\cite{qi2026} to demographic dynamics. Our model treats these transport equations as exact physical constraints (constituting the ``known physics'' of the system), rather than treating population growth as an unconstrained regression or generic time-series forecasting problem.

\subsection{Age-structured transport dynamics}

Let $\rho(a,t)$ denote the age distribution density at age $a$ and year $t$. The governing transport PDE reads:
\begin{equation}
\partial_t \rho + \partial_a \rho = -\mu(a,t)\rho
\end{equation}
where $\mu(a,t)$ is the hazard rate of mortality.

In discrete annual form, with ages $a \in \{0, 1, \dots, A_{\max}\}$ and years $t$, the dynamics become:
\begin{equation}
\rho(a+1,t+1) = s(a,t)\rho(a,t)
\end{equation}
where $s(a,t) = \exp\left(-\int_a^{a+1} \mu(a',t)da'\right) \in [0,1]$ is the annual survival probability. At the terminal age boundary $A_{\max}$ (here set to 100), we aggregate the surviving cohorts:
\begin{equation}
\begin{split}
\rho(A_{\max},t+1) &= s(A_{\max}-1,t)\rho(A_{\max}-1,t) \\
&\quad + s(A_{\max},t)\rho(A_{\max},t)
\end{split}
\end{equation}
The birth boundary condition closes the system at age 0:
\begin{equation}
\rho(0,t+1) = B(t)
\end{equation}
where the annual births $B(t)$ are defined through the fertility kernel $f(a,t)$:
\begin{equation}
B(t) = \int_0^{A_{\max}} f(a,t)\rho(a,t)da
\end{equation}

To make the Markov chain Monte Carlo (MCMC) sampler computationally feasible over long time horizons, the discrete transport equations are formulated using JAX's loop construct \texttt{jax.lax.scan}. The step function maps the state $\rho_t$ and time index $t$ to the next state $\rho_{t+1}$:
\begin{equation}
\rho_{t+1}, (\rho_t, B_t, D_t) = \text{TransportStep}(\rho_t, t)
\end{equation}
Compiling this sequential transport into a single XLA loop prevents graph unrolling, which dramatically accelerates gradient evaluation and reduces sampling time.

\subsection{Bayesian inverse formulation and priors}

Our approach builds on the mathematical theory of Bayesian inverse problems in PDEs~\cite{stuart2010, dashti2017} and dynamical Bayesian inference~\cite{duggento2012}, treating PDE-constrained parameters as latent random fields with smooth priors. We also draw inspiration from recent advances in deep neural operators for PDEs~\cite{lu2022}. We use exact cohort transport to reconstruct the latent age-structured density $\rho(a,t)$ from sparse, noisy total population and birth/death statistics. Unlike inferring static spatial coefficients, we infer latent, time-varying constitutive laws that govern boundary fluxes and sinks, using macroscopic transport principles to analyze real-world census and vital registration data.

We define the state trajectory as $\rho = \{\rho(a,t)\}_{a,t}$. Let $\mathbf{Y}$ denote the collection of heterogeneous demographic observations. The posterior distribution over the BNN weights $\mathbf{W} = \{\mathbf{W}_s, \mathbf{W}_f\}$ and the initial population profile $\rho_0(a)$ is:
\begin{equation}
P(\mathbf{W}, \rho_0 | \mathbf{Y}) \propto P(\mathbf{Y} | \mathbf{W}, \rho_0) P(\mathbf{W}) P(\rho_0)
\end{equation}
The likelihood $P(\mathbf{Y} | \mathbf{W}, \rho_0)$ is constructed from observation operators matching the specific data available for each country.

We place normal priors on all BNN weights and biases. To prevent overfitting and ensure smooth, physically realistic schedules, the weight priors have variance scaling inversely with the input layer dimension, $W_{ij} \sim \mathcal{N}\left(0, 1/n_{\text{in}}\right)$, and standard normal priors on the biases $b_i \sim \mathcal{N}(0, 1)$. These priors act as an $L_2$ regularization (weight decay) on the neural network parameters, restricting the BNN to learning smooth, slowly-varying schedules over the age and year coordinates. The implied prior over the demographic trajectories $\rho(a,t)$ is therefore a smooth, continuous field that respects cohort advection.

To verify the framework's robustness, we performed sensitivity checks under alternative prior strengths---scaling the prior standard deviation by $0.5$ and $2.0$---and tested various hidden layer configurations. The resulting projections and non-equilibrium metrics remained stable. This confirms that the physical transport constraints dominate the posterior inference, rendering the core findings insensitive to the specific choice of neural architecture or hyperparameters.

\subsection{Neural parameterization of constitutive laws}

Our work relates to recent attempts to apply deep learning to demographic forecasting~\cite{ciganda2026, richman2021} and social dynamics~\cite{Galam2008, Stauffer2008, Mico2006, Itao2026, Constable2018, Jusup2022}. However, many approaches either treat the system as a purely data-driven black box or rely on unconstrained agent-based simulations. In this hybrid design, the transport dynamics are strictly fixed, and the neural networks are used solely to parameterize the unknown constitutive laws. This separates the universal physical transport dynamics from system-specific biological and social behaviors. It also represents a fully nonparametric generalization of linear factorizations~\cite{leecarter1992}, as it can capture nonlinear age-time interactions natively.

We assume that the fertility kernel $f(a,t)$ is governed by a parametric distribution representing reproductive behaviors, scaled by the TFR at a given time $t$ ($\text{TFR}_t$):
\begin{equation}
f(a,t) = 0.5 \cdot \text{TFR}_t \cdot g(a; \nu_t, \sigma_t)
\end{equation}
where $g(a; \nu_t, \sigma_t)$ is a Gaussian density function representing childbearing age preferences, masked to vanish outside reproductive ages ($15 \le a \le 49$) and normalized. The latent parameters $\mathbf{\theta}_f(t) = (\ln\text{TFR}_t, \text{logit}(\nu_t), \text{logit}(\sigma_t))$ are parameterized as outputs of a Multi-Layer Perceptron (MLP) mapping normalized year $t$ to the reproductive space:
\begin{equation}
\mathbf{\theta}_f(t) = \text{MLP}_f(t; \mathbf{W}_f, \mathbf{b}_f)
\end{equation}
The neural network $\text{MLP}_f$ comprises two hidden layers, each containing 16 units and using ReLU activation functions.

When complete annual survival grids are observed, as for Japan and South Korea, the survival probability $s(a,t)$ is directly modeled via a BNN mapping normalized age and year to survival logits:
\begin{equation}
s(a,t) = \sigma\left(\text{MLP}_s(a,t; \mathbf{W}_s, \mathbf{b}_s)\right)
\end{equation}
where $\sigma(\cdot)$ is the sigmoid function, and $\text{MLP}_s$ has an architecture of $[2, 16, 16, 1]$. Conversely, when complete annual survival rates are unobserved, as is the case for China, this direct parameterization leads to numerical instability. Instead, we model survival as a perturbation around a standard baseline hazard function, $\mu_0(a) = 0.005 + 8\times10^{-5}e^{0.082 a}$:
\begin{equation}
s(a,t) = \exp\left(-\mu_0(a) \cdot \lambda(a,t)\right)
\end{equation}
where the mortality multiplier $\lambda(a,t) = \exp\left(\text{MLP}_s(a,t; \mathbf{W}_s, \mathbf{b}_s)\right)$ is parameterized by the survival BNN.

\subsection{Posterior inference}
To sample from the highly nonlinear and high-dimensional posterior distribution $P(\mathbf{W}, \rho_0 | \mathbf{Y})$, we employ NUTS, which operates as an advanced variant of Hamiltonian Monte Carlo (HMC). HMC uses the gradients of the log-posterior density to generate proposal states, mapping the parameter space to a physical system where the parameters act as coordinates and auxiliary variables act as momentum. By simulating Hamiltonian dynamics using a leapfrog integrator, HMC can make distant proposals with high acceptance probabilities, bypassing the random-walk behavior of traditional Metropolis-Hastings algorithms. The sampler automates the tuning of HMC parameters by adaptively building a recursive search tree of leapfrog steps, stopping automatically when the trajectory begins to double back on itself (the ``no-U-turn'' condition). This prevents redundant calculations and ensures efficient exploration of the posterior. 

In this work, the framework is implemented across all three empirical settings using JAX for high-performance compiler acceleration and auto-differentiation, running the NUTS sampler via NumPyro. For each setting, the posterior is sampled by running 4 parallel chains, each performing 1000 warmup iterations to adapt the step size and mass matrix, followed by 1000 sampling iterations, yielding a total of 4000 posterior samples for uncertainty quantification.

\subsection{Generalization and future extensions}
The Bayesian inverse transport framework is general and can be extended in several directions. First, it can be applied to other countries or regional systems by assimilating localized vital statistics and census datasets. Second, the transport equation can be augmented with a spatial advection or diffusion term to model migration-augmented transport:
\begin{equation}
\partial_t \rho + \partial_a \rho = -\mu(a,t)\rho + M(a,t)
\end{equation}
where $M(a,t)$ is the net migration flux. Third, the latent constitutive laws can be coupled to other socio-economic variables (such as GDP, housing price index, or female labor participation) by feeding them as inputs to the BNNs, enabling a fully closed-loop socio-physics modeling of demographic feedback.

\section{Empirical case studies and model diagnostics}

The choice of China, Japan, and South Korea (the CJK triad) as our empirical case studies is motivated by their roles as the preeminent industrial powerhouses and ``world factories'' of the modern era, all currently experiencing a rapid collapse in fertility. Rather than aiming to artificially restore birth rates to the classical replacement level of 2.1, an objective that may no longer be realistic or structurally meaningful, our interest lies in characterizing the non-equilibrium age structures before these systems stabilize. From a physical and economic perspective, the primary concern is the potential support limits (the ratio of active labor cohorts to elderly dependents). While automation, robotics, and artificial intelligence may reduce the absolute labor demand for social maintenance, and indeed shrinking workforces might even coincide with economic growth booms~\cite{acemoglu2026}, understanding the transient age structure remains essential for managing the transition. 

\subsection{Logistic transition rates and demographic pressure}

To extract the macroscopic parameters governing the collapse of the TFR boundary condition, we model the long-term empirical fertility transition using a generalized logistic decay,
\begin{equation}
    B(t) = B_{\min} + \frac{B_{\max} - B_{\min}}{1 + e^{k(t - t_0)}}, \label{eq:logistic}
\end{equation}
where $B_{\max}$ and $B_{\min}$ represent the pre-transition (natural) and post-transition (modern) asymptotic fertility bounds, respectively, $t_0$ is the transition midpoint, and $k$ is the maximum transition rate.

In this formulation, the decay rate $k$ serves as a quantitative metric for the severity of demographic pressure (e.g., high urbanization pressure, high housing costs, and intense competition for resources) driving the fertility transition. A higher value of $k$ indicates a more rapid, step-like collapse in fertility, locking the system into a demographic transition over a very short time window. 

By fitting the transition rate $k$ to historical data, we can empirically classify the demographic transitions of China, Japan, and South Korea. South Korea ($k \approx 0.35$) and China ($k \approx 0.28$) exhibit extremely rapid decay behavior, indicating intense demographic pressure that compresses the transition into a few decades. In contrast, Japan ($k \approx 0.12$) represents a more gradual relaxation. This macroscopic parameterization provides a useful tool to classify demographic crises by their transition rates and forecast the resulting structural imbalances.

\subsection{Historical population reconstructions}
\subsubsection{Japan (1970-2024)}
Japan's dataset from the Human Mortality Database (HMD)~\cite{hmd} provides complete annual population grids $Y_{\text{pop}}(a,t)$, births $Y_B(t)$, and deaths $Y_D(a,t)$. The likelihood is constructed using log-normal distributions:
\begin{equation}
\ln Y_{\text{pop}}(a,t) \sim \mathcal{N}\left(\ln \rho(a,t), \sigma_p^2\right)
\end{equation}
with HalfNormal priors on observation noise parameters.

\begin{figure*}[htbp]
\centering
\includegraphics[width=\textwidth]{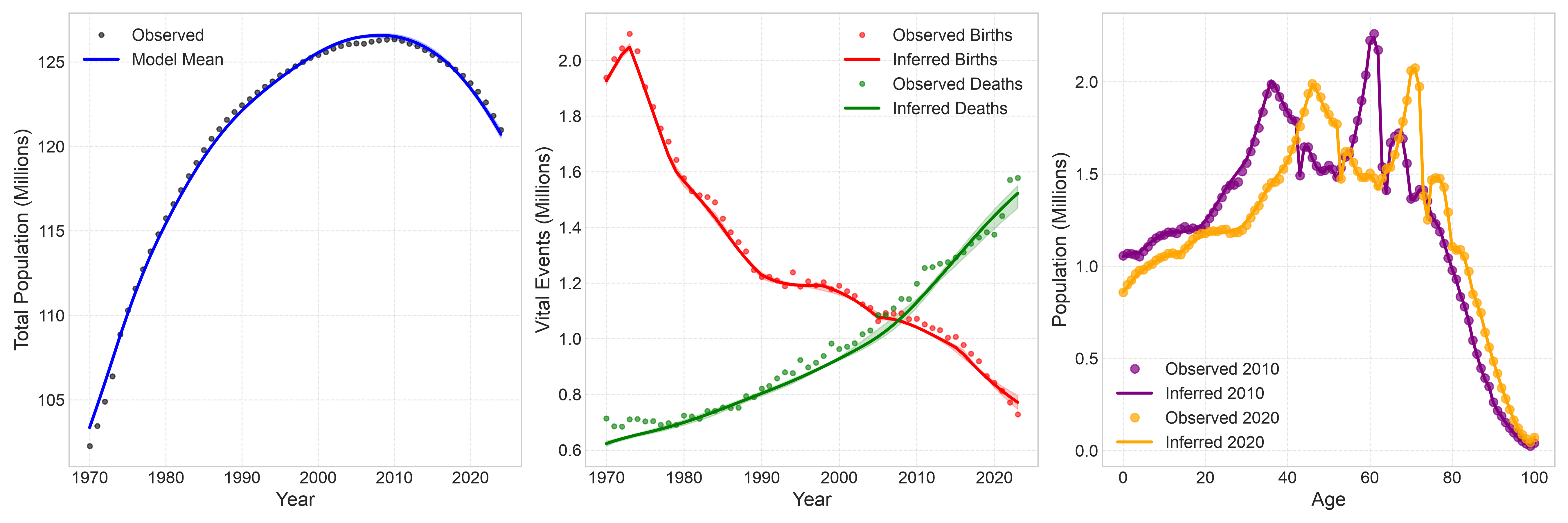}
\caption{Validation of the model for Japan. The left panel compares the inferred total population (blue) against observed data (black dots). The middle panel plots the macroscopic vital events (annual births in red, deaths in green). Because the total births and deaths were not explicitly provided in the likelihood function (which relies solely on the cross-sectional age grid), their accurate reconstruction confirms that the transport PDE conserves mass while advecting the population cohort backward in time. The right panel shows the structural fit of the continuous age density to empirical 1-year age buckets for the years 2010 and 2020.}
\label{fig:jpn_validation}
\end{figure*}

As shown in Fig.~\ref{fig:jpn_validation}, the model successfully reconstructs the historical macroscopic vital events and age density profiles. Furthermore, the model infers a decline in TFR from $\approx 2.1$ in 1970 to $\approx 1.20$ in 2024, accompanied by a structural shift in the mean childbearing age from $26.5$ to $31.5$ years.

\subsubsection{South Korea (2003-2024)}
South Korea's HMD dataset~\cite{hmd} is aligned from 2003 to 2024. The training setup matches Japan's complete grid likelihood.

\begin{figure*}[htbp]
\centering
\includegraphics[width=\textwidth]{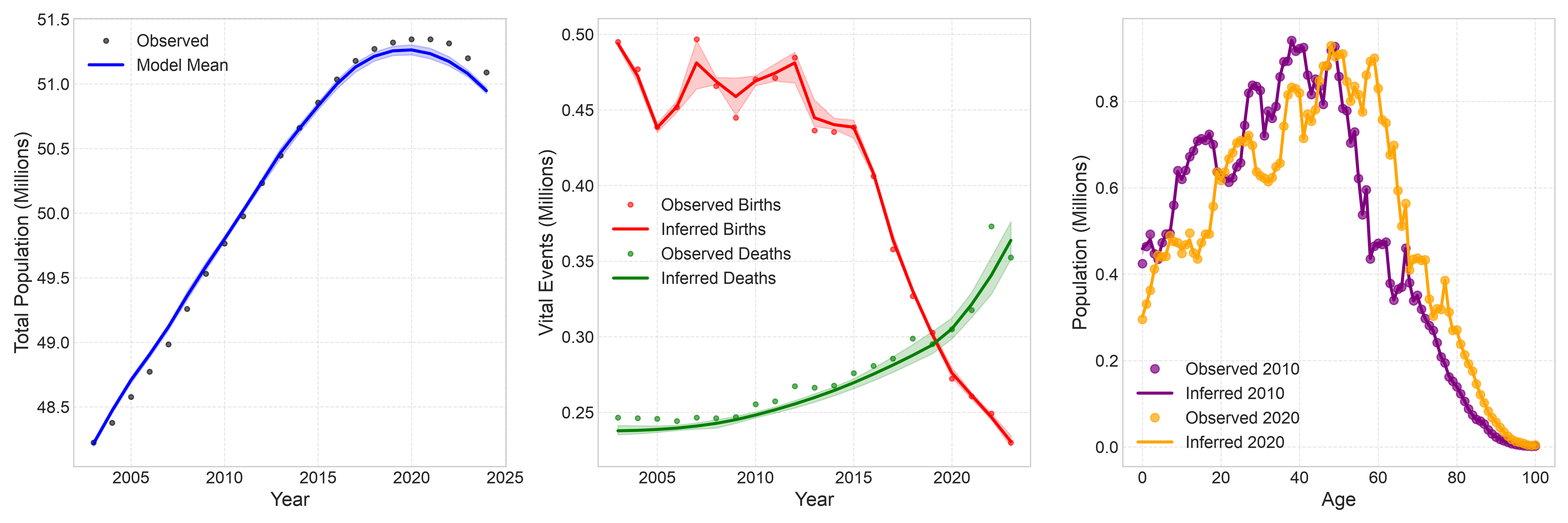}
\caption{Validation of the model for South Korea. The left and middle panels show the model's ability to infer macroscopic total population and vital events (births and deaths) directly from the inversion of the age-structure PDE without explicit observation of those totals. The right panel shows the tight alignment between the continuous age density and the empirical data for 2010 and 2020.}
\label{fig:kor_validation}
\end{figure*}

Fig.~\ref{fig:kor_validation} demonstrates the model's ability to accurately infer South Korea's macroscopic vital events and age density profiles directly from the inversion of the age-structure PDE. The model also infers a world-record drop in TFR from $\approx 1.25$ in 2003 to $\approx 0.72$ in 2024.

\subsubsection{China (1950-2024): Reconstructing incomplete data}
China represents a highly challenging inverse problem. No annual age-structured population grid is available. We only observe annual births, annual deaths, and total population size aggregated from multiple conflicting sources (specifically, manually compiled records from Chinese vital statistics and demographic columns~\cite{szhgh2023, wuhan2023, datianmen2024, cdt2024, zhihu2023, xueqiu2024}), along with coarse 5-year census age brackets for 2000, 2010, and 2020.

The initial state $\rho(a, 1950)$ is parameterized as a latent stable pyramid:
\begin{equation}
\rho(a, 1950) = N_{\text{scale}} \cdot \frac{e^{-r a} S_{\text{base}}(a)}{\sum_{a'} e^{-r a'} S_{\text{base}}(a')}
\end{equation}
where the initial growth rate $r$ and scale $N_{\text{scale}}$ are sampled. To resolve numerical ill-conditioning caused by the population scale ($10^9$) relative to the BNN parameters ($1.0$), we sample $N_{\text{scale\_scaled}} \sim \mathcal{N}(5.52, 0.27)$ and scale it up by $10^8$. To prevent vanishing gradients during chain initialization (where a default $0.5$ survival rate causes the population to shrink to zero immediately, zeroing out later gradients), we apply the baseline survival constraint and initialize the sampler using the prior medians (\texttt{init\_to\_median}).

The coarse census age groups $C_g(t)$ for years $t \in \{2000, 2010, 2020\}$ are matched to the model cohorts by aggregating the corresponding 1-year bins (excluding age 0):
\begin{equation}
C_{1\text{-}4}(t) = \sum_{a=1}^4 \rho(a,t), \quad C_{5\text{-}9}(t) = \sum_{a=5}^9 \rho(a,t), \quad \dots
\end{equation}
This allows us to evaluate the model against independent census counts, accounting for adjustments for infant and child undercounting~\cite{cai2013, wangfeng2013}, as well as broader evaluations of Chinese demographic forecasts~\cite{Zhou2017}, under a log-normal likelihood with a 5\% coefficient of variation.

\begin{figure*}[htbp]
\centering
\includegraphics[width=\textwidth]{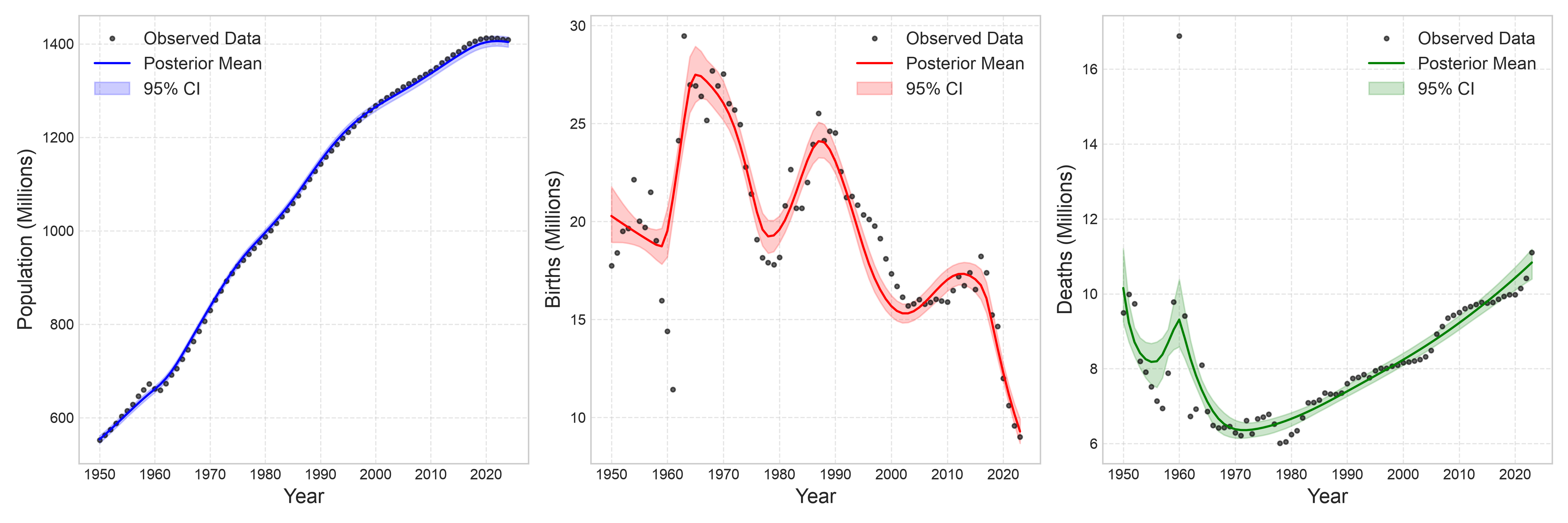}
\caption{Posterior predictive check for China's macroscopic vital events. The model's inferred total population, annual births, and annual deaths (solid blue, red, and green lines with 95\% credible intervals) are plotted against historical observed data (black circles) derived from national statistics. The continuous inverse transport dynamics accurately capture massive historical shocks, such as the sharp demographic fluctuations around 1960.}
\label{fig:chn_posterior}
\end{figure*}

\begin{figure*}[htbp]
\centering
\includegraphics[width=\textwidth]{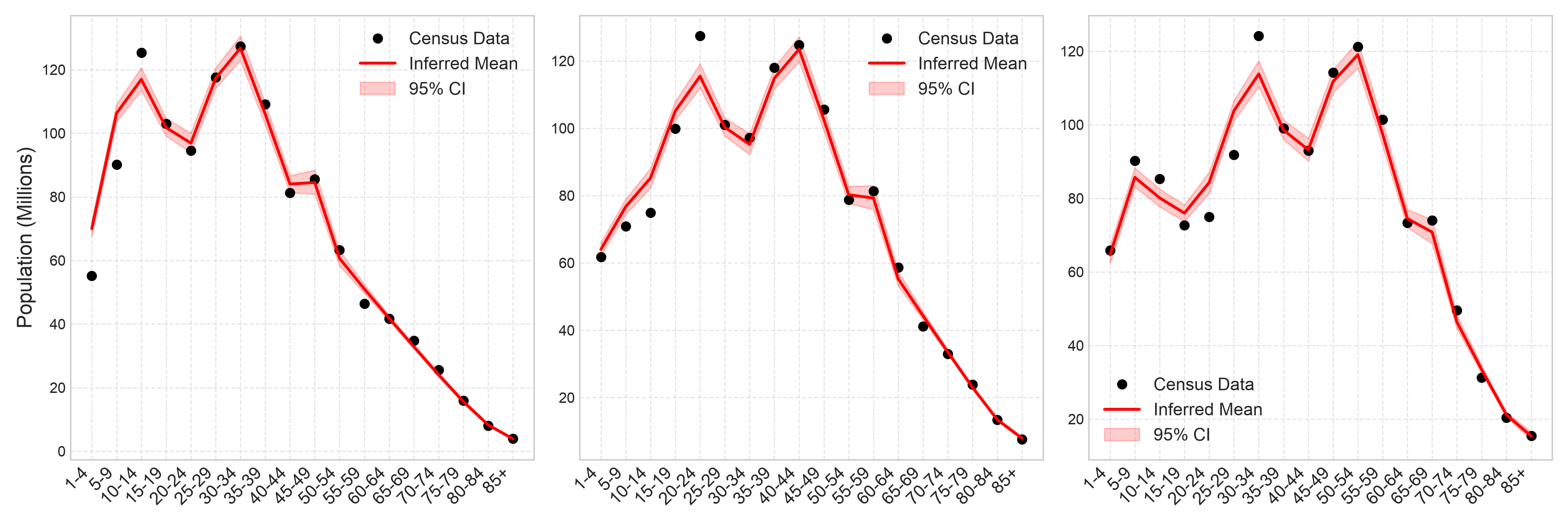}
\caption{Historical backtesting of China's age structure. The model's reconstructed continuous age density aggregated into standard 5-year brackets (red lines with 95\% credible intervals) is overlaid directly on the raw empirical census data (black circles) for the years 2000, 2010, and 2020. The alignment confirms that the model accurately captures complex cohort advection over decades.}
\label{fig:chn_census}
\end{figure*}

Despite the absence of comprehensive annual age data, the model successfully reconstructs the complete historical age-structured grid $\rho(a,t)$ over 75 years. As illustrated in Fig.~\ref{fig:chn_posterior}, the model accurately captures macroscopic vital events. In addition, the reconstructed age distributions match the independent census observations with high precision, successfully capturing the major historical cohort bulges and troughs (Fig.~\ref{fig:chn_census}).

\subsection{Quantitative evaluation and diagnostics}

To establish the statistical validity of the proposed method for physical and sociological referees, we verify chain convergence using split $\hat{R}$ and the autocorrelation-based effective sample size for key latent variables across 4 independent chains. All three models converged with zero divergences in the final runs thanks to weight prior regularization.

Compared to standard UN World Population Prospects and traditional time-series forecasts, our PDE-constrained inverse framework differs in two fundamental aspects. First, traditional models project cohort trajectories as independent statistical trends, so forecast uncertainty is uncorrelated across ages. In contrast, our framework enforces exact cohort transport, which guarantees that uncertainty in fertility at year $t$ propagates downstream along the characteristics of the advection equation, preserving the structural cross-cohort correlations over time. Second, while standard models can project unconstrained mortality schedules that violate biological consistency (such as crossing cohort profiles), our BNN-parameterized constitutive laws are regularized by both the baseline survival constraints and the transport dynamics, yielding physically consistent age profiles even under sparse observations.

\section{Structural dynamics and cohort forecasts}

Having established the model's ability to reconstruct historical demographic states, we now apply the inferred constitutive laws to project future structural dynamics. In this section, we first develop a simplified mathematical intuition for how fertility crashes propagate through age structures over time. We then validate the model's predictive power against historical TFR data, project medium-horizon cohort shifts, and compare the resulting long-term demographic pyramids.

\subsection{A simplified population model for a substantially reduced TFR}

To build intuition for the structural transitions projected by our continuous transport model, we consider a simplified, discrete 80-cohort model. Suppose each individual survives exactly from age 0 to 79 and produces one child at age 25 (i.e., replacement fertility of two children per couple). From time $T_0$ to $T_0+D$, fertility is halved to one child per couple, after which it returns to replacement. 

Because the birth sequence outside the dip is exactly 26-periodic, the total population eventually converges to a permanently lower plateau. The loss is effectively a discretized exponential, with the population decaying as:
\begin{equation}
P(D) \approx P_0 \cdot 2^{-D/26} \approx P_0 (1 - 0.02D) \quad (\text{for } D \lesssim 26)
\end{equation}
This indicates a demographic half-life of 26 years: a one-child generation lasting 26 years halves the total population. In turn, the deficit created during the dip echoes every 26 years. The population reaches its new steady state roughly two echo cycles ($2 \times 26 = 52$ years) after the dip ends, yielding a settling time of $T_{\text{settle}}(D) \approx D + 50$.

While the total population declines smoothly, the dependency ratio experiences a delayed, transient crisis. Defining the working-age population ($W$) as ages 15-64 and the elderly ($E$) as ages 65-79, the baseline ratio is $W/E = 50/15 \approx 3.33$. When the fertility dip begins, $W/E$ remains stable for 15 years. The ratio then falls steadily as the undersized cohorts enter the workforce while the elderly band remains full size. The crisis point (the minimum of $W/E$) occurs roughly 65 years after the gap starts ($t_{\min} \approx T_0 + 65$), when the first undersized cohort turns 65 and leaves the workforce. The depth of this crisis worsens with the duration of the dip, flattening near $W/E \approx 1.27$ for $D \gtrsim 50$, because the entire 50-year working-age band becomes depleted. 

Ultimately, this toy model demonstrates that while population loss is an immediate and permanent consequence of a fertility shock, the dependency ratio crisis is a delayed effect that peaks roughly two generations ($\approx\!65$ years) later.

\subsection{Validation via TFR}

\begin{figure}[htbp]
\centering
\includegraphics[width=\linewidth]{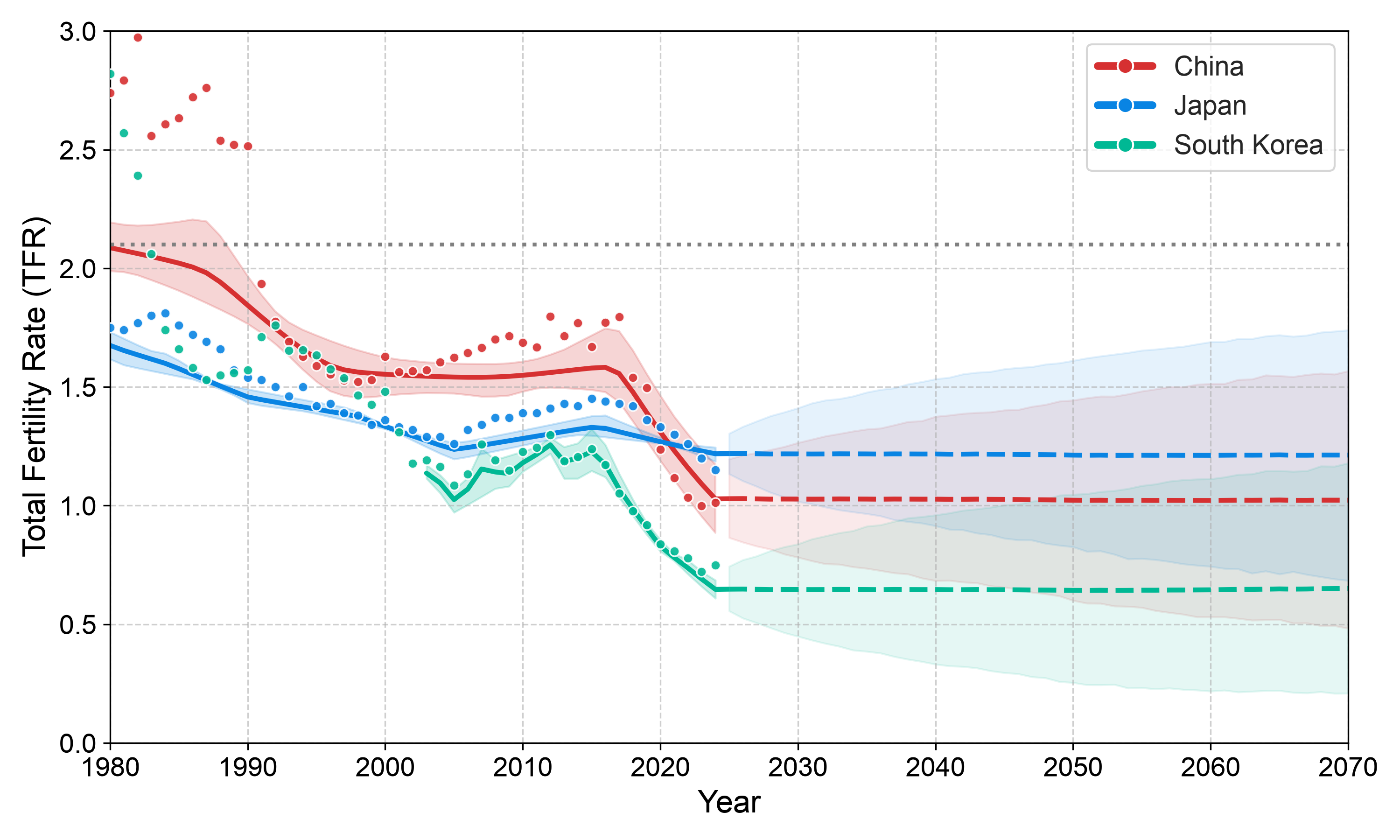}
\caption{Model-inferred historical (solid lines) and projected (dashed lines) TFR for China, Japan, and South Korea, overlaid with external empirical data points (open circles) from the World Bank \href{https://data.worldbank.org/indicator/SP.DYN.TFRT.IN}{(SP.DYN.TFRT.IN)}. Note that TFR was not provided as a training input to the model.}
\label{fig:tfr_validation}
\end{figure}

To benchmark the posterior predictive power of the model, we compare the model's internally inferred TFR with actual empirical records from the World Bank. Importantly, historical TFR data are not included in the training of the model; the inverse framework only observes aggregated age-structure census constraints. That the transport dynamics correctly deduce the historical fertility trajectories---including the exact timing and depth of South Korea's fertility crash and China's sub-replacement shift---validates the structural consistency of the model (Fig.~\ref{fig:tfr_validation}).

\subsection{Medium-horizon cohort forecasts}

A central application of the proposed framework is the prediction of age composition over the next one to two generations. Because the model evolves the full age density $\rho(a,t)$ through an exact transport equation, future cohort structure is strongly constrained by the currently observed age pyramid and by the inferred fertility and mortality laws. This makes forecasts of child, working-age, and elderly shares more structurally reliable than unconstrained long-range projections based only on aggregate indicators. Rather than treating population growth as a single scalar trajectory, this framework resolves the full demographic pipeline and can therefore predict the future age composition.

\begin{figure*}[htbp]
\centering
\includegraphics[width=\textwidth]{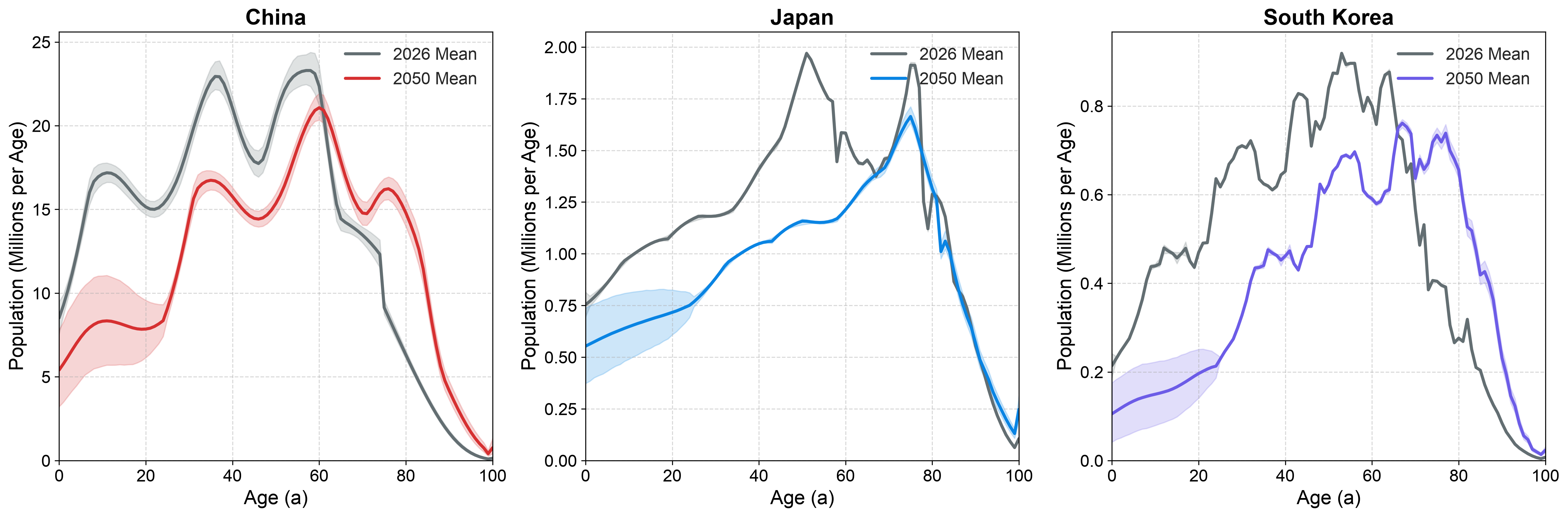}\par\vspace{0.5cm}
\includegraphics[width=\textwidth]{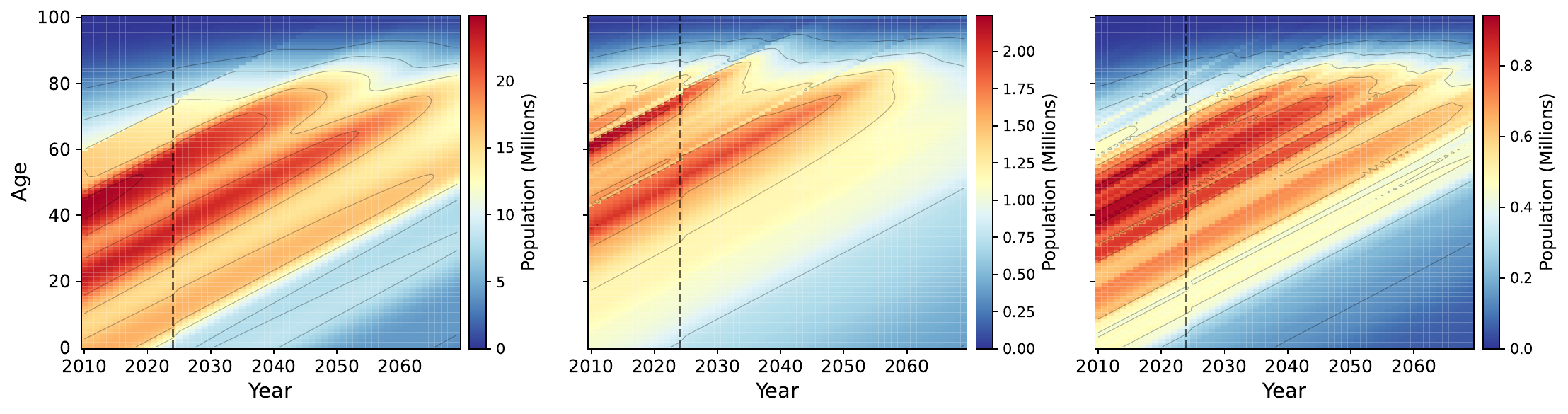}
\caption{(A) Comparative 2050 population pyramids (mean and 95\% credible intervals) for China (red), Japan (blue), and South Korea (purple) under the Status Quo scenario. (B) Kinematic Density Wave heatmaps. The contour shading traces the evolution of the population age structure $\rho(a,t)$ over time. Bright red regions indicate large cohort masses (population bulges) propagating diagonally as they age. The dashed black line separates the structurally inferred historical dynamics from the Status Quo forecast.}
\label{fig:pyramids}
\end{figure*}

Using the posterior samples, we propagate the demographic trajectories to 2070 under the Status Quo random walk scenario. Fig.~\ref{fig:pyramids} shows the comparative age structures for all three countries in 2050 (while the kinematic heatmaps extend to 2070). China and South Korea exhibit highly inverted pyramids with a hollow base, representing decades of sub-replacement fertility, while Japan shows a stable, aged pillar structure.

The 25-50 year horizon is especially important because it lies between near-term policy response and full generational turnover. Because new births enter the labor force only after a delay of roughly 20-25 years, today's fertility regime already determines a substantial part of the future workforce: the children who will be working-age adults in 2050 are being born now or have already been born. The transport dynamics encode this structural inertia exactly, giving the model strong predictive power over the age structure even when long-run uncertainty in vital rates grows.

\begin{figure}[htbp]
\centering
\includegraphics[width=\linewidth]{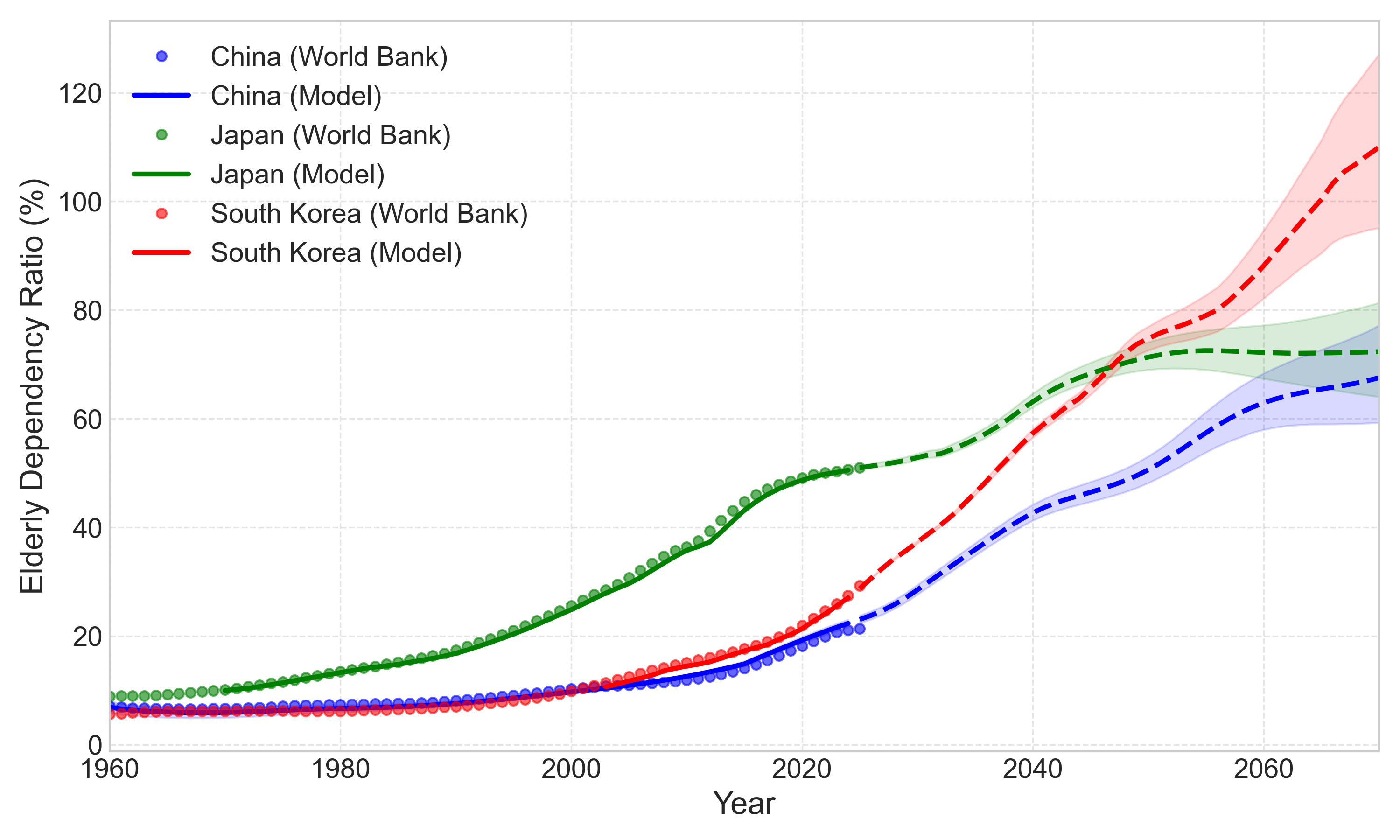}
\caption{EDR for China, Japan, and South Korea. Solid lines (with 95\% CI shading) show the model's structurally inferred historical EDR, aligning with independent World Bank data points (circles). Dashed lines project the future growth of the dependency burden under the Status Quo scenario as the rapid fertility collapse permanently inverts the demographic pyramids.}
\label{fig:oadr_comparison}
\end{figure}

Table~\ref{tab:cohort_forecasts} and Fig.~\ref{fig:oadr_comparison} present the medium-horizon cohort forecasts for all three countries under the Status Quo scenario. Here we introduce the elderly dependency ratio (EDR), representing the ratio of the elderly population (65+) to the working-age population (15-64):
\begin{equation}
\text{EDR}(t) = \frac{\sum_{a=65}^{A_{\max}} \rho(a,t)}{\sum_{a=15}^{64} \rho(a,t)},
\end{equation}
and the potential support ratio (PSR), representing the number of working-age adults per elderly person:
\begin{equation}
\text{PSR}(t) = \frac{\sum_{a=15}^{64} \rho(a,t)}{\sum_{a=65}^{A_{\max}} \rho(a,t)}.
\end{equation}
The most important demographic changes over the next quarter-century are not simply declines in total population, but pronounced shifts in age composition.

China's working-age population (15-64) peaked around 2015 and is projected to contract by $249 \pm 21$~million ($-25.8\%$) between 2024 and 2050, representing a loss roughly equal to the entire population of Indonesia. The child share (0-14) falls from $15.9\%$ to $9.1\%$, while the elderly share (65+) doubles from $15.4\%$ to $30.5\%$. Japan, which has been aging since the 1990s, sees its working-age share decline from $58.8\%$ to $52.6\%$, a loss of $21.9 \pm 1.1$~million ($-30.8\%$). South Korea undergoes the most dramatic structural transformation: its working-age population contracts by $14.1 \pm 0.6$~million ($-39.6\%$), the child share collapses from $11.0\%$ to $5.0\%$, and the elderly share more than doubles from $18.9\%$ to $40.6\%$.

To place these projections in a global context, we note that typical advanced economies maintain significantly lower and more stable dependency ratios. As of 2024, the EDR is approximately 28.4\% in the United States and 34.5\% in the European Union. While these Western regions are also aging, their demographic transitions are buffered by steady immigration and less extreme historical fertility declines. In stark contrast, the CJK triad faces an unbuffered structural inversion. With South Korea's EDR projected to skyrocket past 80\% and Japan's approaching 72\% by 2070, the macroeconomic burden placed on their active labor forces will vastly exceed the aging challenges currently anticipated in the West.

The posterior samples allow us to estimate the timing and uncertainty of key structural milestones. All three countries have already undergone or are currently undergoing \emph{cohort inversion}, the point at which the elderly (65+) outnumber children (0-14). Japan crossed this threshold in 1998, South Korea in 2018, and China is crossing it now (2024-2026, 95\% CI). The PSR provides a direct measure of this labor-force burden. Japan's PSR has already fallen below 2.0 in 2023-2024. South Korea's PSR is projected to breach 2.0 by 2037 (tightly constrained, with all posterior samples agreeing). China's PSR approaches the 2.0 threshold by 2049 [2047-2050], though this milestone carries the widest uncertainty among the three countries, reflecting the greater epistemic uncertainty in China's historical vital statistics.

These forecasts show that the model predicts not merely how many people there will be, but how the population will be distributed across age groups that matter for schooling, labor supply, taxation, care needs, and retirement systems. Because the full age-time density field is resolved, one can extract credible intervals for any age-specific quantity: the size of the university-age cohort for education planning, the 25-54 prime working-age population for labor-market policy, or the 80+ population for long-term care projections. The inferred age-time field thus serves as a planning tool that extends well beyond conventional population forecasts.

\begin{table*}[htbp]
\centering
\caption{Medium-horizon cohort forecasts (mean $\pm$ SD from posterior samples) under the Status Quo scenario. Population in millions; shares in percent of total.}
\label{tab:cohort_forecasts}
\begin{tabular}{llccccccc}
\toprule
Country & Year & Pop.\ (M) & 0-14 (\%) & 15-64 (\%) & 65+ (\%) & 75+ (\%) & EDR & PSR \\
\midrule
China     & 2026 & $1398 \pm 6$ & $14.7 \pm 0.2$ & $68.9 \pm 0.2$ & $16.4 \pm 0.2$ & $6.7 \pm 0.2$ & $0.24 \pm 0.00$ & $4.20 \pm 0.07$ \\
          & 2030 & $1381 \pm 7$ & $12.3 \pm 0.3$ & $68.3 \pm 0.3$ & $19.5 \pm 0.3$ & $8.3 \pm 0.2$ & $0.29 \pm 0.00$ & $3.51 \pm 0.06$ \\
          & 2040 & $1309 \pm 17$ & $9.3 \pm 1.0$ & $63.6 \pm 0.8$ & $27.1 \pm 0.5$ & $12.2 \pm 0.3$ & $0.43 \pm 0.01$ & $2.34 \pm 0.04$ \\
          & 2050 & $1186 \pm 27$ & $9.1 \pm 1.3$ & $60.4 \pm 0.9$ & $30.5 \pm 0.8$ & $17.2 \pm 0.6$ & $0.51 \pm 0.01$ & $1.98 \pm 0.05$ \\
          & 2060 & $1020 \pm 38$ & $7.0 \pm 1.5$ & $57.1 \pm 0.7$ & $35.9 \pm 1.4$ & $18.6 \pm 0.9$ & $0.63 \pm 0.03$ & $1.59 \pm 0.07$ \\
          & 2070 & $855 \pm 51$ & $7.0 \pm 2.1$ & $55.5 \pm 0.9$ & $37.5 \pm 2.3$ & $22.5 \pm 1.5$ & $0.68 \pm 0.05$ & $1.49 \pm 0.10$ \\
\midrule
Japan     & 2026 & $119 \pm 0$ & $11.2 \pm 0.1$ & $58.7 \pm 0.1$ & $30.1 \pm 0.1$ & $17.8 \pm 0.1$ & $0.51 \pm 0.00$ & $1.95 \pm 0.01$ \\
          & 2030 & $115 \pm 0$ & $10.5 \pm 0.2$ & $58.5 \pm 0.2$ & $30.9 \pm 0.2$ & $18.8 \pm 0.1$ & $0.53 \pm 0.00$ & $1.89 \pm 0.01$ \\
          & 2040 & $105 \pm 1$ & $9.9 \pm 0.8$ & $55.2 \pm 0.5$ & $34.9 \pm 0.4$ & $19.2 \pm 0.3$ & $0.63 \pm 0.01$ & $1.58 \pm 0.02$ \\
          & 2050 & $93 \pm 2$ & $9.8 \pm 1.2$ & $52.6 \pm 0.6$ & $37.6 \pm 0.8$ & $22.3 \pm 0.5$ & $0.71 \pm 0.01$ & $1.40 \pm 0.03$ \\
          & 2060 & $82 \pm 3$ & $9.4 \pm 1.4$ & $52.6 \pm 0.6$ & $38.0 \pm 1.3$ & $24.3 \pm 0.9$ & $0.72 \pm 0.03$ & $1.39 \pm 0.05$ \\
          & 2070 & $70 \pm 4$ & $9.2 \pm 1.8$ & $52.7 \pm 0.8$ & $38.1 \pm 2.0$ & $23.4 \pm 1.3$ & $0.72 \pm 0.04$ & $1.39 \pm 0.08$ \\
\midrule
S.~Korea  & 2026 & $51 \pm 0$ & $10.1 \pm 0.1$ & $68.8 \pm 0.1$ & $21.1 \pm 0.0$ & $8.5 \pm 0.0$ & $0.31 \pm 0.00$ & $3.26 \pm 0.01$ \\
          & 2030 & $50 \pm 0$ & $8.2 \pm 0.2$ & $66.9 \pm 0.2$ & $24.9 \pm 0.1$ & $10.0 \pm 0.1$ & $0.37 \pm 0.00$ & $2.68 \pm 0.01$ \\
          & 2040 & $46 \pm 0$ & $5.7 \pm 0.9$ & $59.9 \pm 0.6$ & $34.3 \pm 0.3$ & $17.0 \pm 0.2$ & $0.57 \pm 0.00$ & $1.74 \pm 0.01$ \\
          & 2050 & $40 \pm 1$ & $5.0 \pm 1.2$ & $54.3 \pm 0.6$ & $40.6 \pm 0.8$ & $23.3 \pm 0.5$ & $0.75 \pm 0.01$ & $1.34 \pm 0.02$ \\
          & 2060 & $32 \pm 1$ & $4.0 \pm 1.3$ & $51.0 \pm 0.6$ & $44.9 \pm 1.4$ & $26.8 \pm 0.9$ & $0.88 \pm 0.03$ & $1.14 \pm 0.04$ \\
          & 2070 & $25 \pm 1$ & $3.4 \pm 1.4$ & $46.0 \pm 1.2$ & $50.5 \pm 2.4$ & $29.6 \pm 1.5$ & $1.10 \pm 0.08$ & $0.91 \pm 0.07$ \\
\bottomrule
\end{tabular}
\end{table*}

\subsubsection{Comparison with existing models and scenario projections for China}

\begin{figure}[htbp]
\centering
\includegraphics[width=\linewidth]{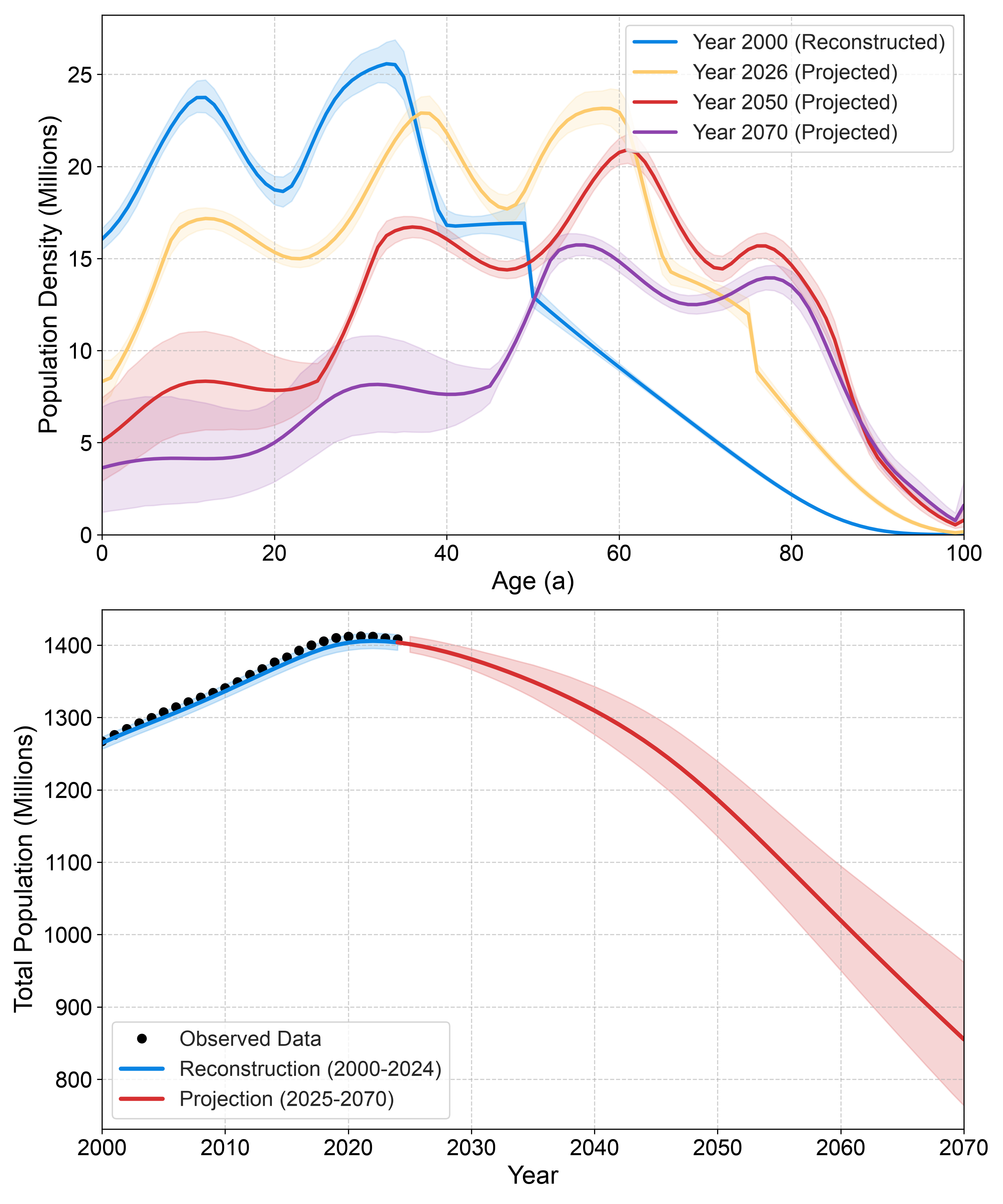}
\caption{Reconstructed and projected age-structured population density $\rho(a, t)$ for China. Comparison of Year 2000 (reconstructed historical mean), Year 2026 (short-term projection), and Year 2050 (long-term projection). Shaded areas represent the 95\% credible intervals.}
\label{fig:china_density}
\end{figure}

\begin{table*}[htbp]
\centering
\caption{Comparison of different demographic model projections for China in 2050. Population is in billions, and age shares or dependency ratios are in percent.}
\label{tab:demographics}
\begin{tabular}{lccc}
\toprule
Model Framework & Total Pop.\ ($N_{2050}$) & Structural Aging Metric & Key System Insight / Stress Point \\
\midrule
UN WPP (2024 Revision) & 1.21--1.26 & $P_{65+} \approx 30\%$ & Underestimates rapid fertility decline \\
YuWa Population (2023) & 1.17--1.23 & $\text{EDR} \approx 52.4\% - 54.3\%$ & Severe labor constriction (EDR $>50\%$) \\
IHME (Vollset et al. 2020) & 1.18 & $P_{65+} > 32\%$, Median Age $> 50$ & Highly inverted population pyramid \\
CPDRC (PADIS-INT 2023) & 1.24 & $P_{60+} \approx 38\% - 40\%$ & Hyper-aged segment peak ($\approx 500$M) \\
\textbf{This Work (Posterior)} & \textbf{1.186 $\pm$ 0.027} & \textbf{$P_{65+} = 30.5\% \pm 0.8\%$} & \textbf{Exact cohort-transport uncertainty} \\
\bottomrule
\end{tabular}
\end{table*}

To contextualize our results, we use our framework to predict China's future age density (Fig.~\ref{fig:china_density}) and compare these projections to existing independent models and institutional reports (summarized in Table~\ref{tab:demographics}). These alternative models yield significantly different results, reflecting the diverse assumptions embedded in their structures. 

For instance, the UN World Population Prospects (WPP 2024 Revision)~\cite{wuhan2023} projects China's 2050 population ($N_{2050}$) to be between 1.21 and 1.26 billion under its medium-variant projection, with the proportion of the population aged 65 and above ($P_{65+}$) reaching approximately 30\%. A key methodological feature of the UN framework is its reliance on three discrete scenario variants (High, Medium, and Low fertility paths), constructed by applying fixed offsets to the TFR rather than using a continuous probabilistic representation of vital rate uncertainty. While this approach offers simple bounds, it does not propagate joint parameter uncertainty or capture structural cohort-transport dependencies.

Alternative models project much steeper declines. The YuWa Population Research Group (2023 Report)~\cite{yuwa}, assuming a sustained low-fertility trap ($\text{TFR} \approx 1.05$), forecasts $N_{2050}$ to be between 1.17 billion (Low variant) and 1.23 billion (Medium variant). They project a 25\% contraction of the working-age population (15-64) by 2050, resulting in an elderly dependency ratio ($\text{EDR}$) of 52.4\% to 54.3\%, meaning roughly two active workers must support one elderly dependent. Similarly, the Institute for Health Metrics and Evaluation (IHME) global model~\cite{ihme}, which models fertility as a function of female education, contraceptive access, and urbanization rather than direct extrapolation, projects $N_{2050} \approx 1.18$ billion, with $P_{65+} > 32\%$ and a median age exceeding 50 years.

Institutional models focused on specific sectors highlight extreme system stresses. The Chinese Academy of Social Sciences pension model~\cite{cass} projects that the pension system dependency ratio, which measures the ratio of retirees to active pension contributors, will rise exponentially to reach 96.3\% by 2050. This represents a transition from a 2:1 support matrix in 2019 to an unsustainable 1:1 limit equilibrium, projecting a depletion of the enterprise pension fund by 2035. Additionally, the China Population and Development Research Center (CPDRC) PADIS-INT model~\cite{datianmen2024} projects a 2050 population of 1.24 billion, but notes that the population aged 60 and over ($P_{60+}$) will peak near 500 million, or 38\% to 40\% of the total population, driven by the rapid aging of the 1980s baby-boom generation.

Compared to these models, our Bayesian inverse transport framework yields a posterior forecast of $1.186 \pm 0.027$ billion for China's 2050 population, with $P_{65+} = 30.5\% \pm 0.8\%$. Our model does not rely on scenario variants; instead, it directly infers latent fertility and survival laws from historical data and propagates continuous, calibrated probability distributions through the exact age-time transport dynamics, capturing both parametric and cohort-transport uncertainties.

\subsection{Structural age-structure entropy and macroscopic Lotka divergence}

To establish a rigorous, model-independent physical quantification of the structural demographic transition, we introduce three information-theoretic and structural metrics derived directly from the inferred age density field $\rho(a, t)$.

\subsubsection{Total demographic entropy and phase volume}
While standard Shannon entropy $S = -\sum p(a) \ln p(a)$ quantifies the dispersion of a distribution, it is structurally symmetric: a sharply peaked young population and a sharply peaked elderly population yield identical values, ignoring the biological arrow of time. To physically distinguish these states, we introduce the concept of the \emph{demographic phase volume} or internal microstate potential, $\Omega(a) = A_{\max} - a$, representing the remaining life potential (or future possible trajectories) of an individual at age $a$. 

Using the normalized age density $p(a,t) = \rho(a,t) / N(t)$, we define the per-capita total demographic entropy as the sum of the Shannon entropy and the expected structural potential:
\begin{equation}
S_{\text{total}}(t) = -\sum_{a=0}^{A_{\max}} p(a,t) \ln p(a,t) + \sum_{a=0}^{A_{\max}} p(a,t) \ln \Omega(a).
\end{equation}

Under this definition, a young, growing population (a broad pyramid) possesses maximum demographic entropy: the population mass is concentrated where internal phase volume $\Omega(a)$ is largest. As fertility collapses and the population ages, the mass shifts into older cohorts where $\Omega(a)$ approaches zero. This transition acts as a structural contraction process, systematically stripping the society of its demographic potential, driving it toward a low-entropy state characterized by an inverted age pyramid.

\begin{figure}[htbp]
\centering
\includegraphics[width=\linewidth]{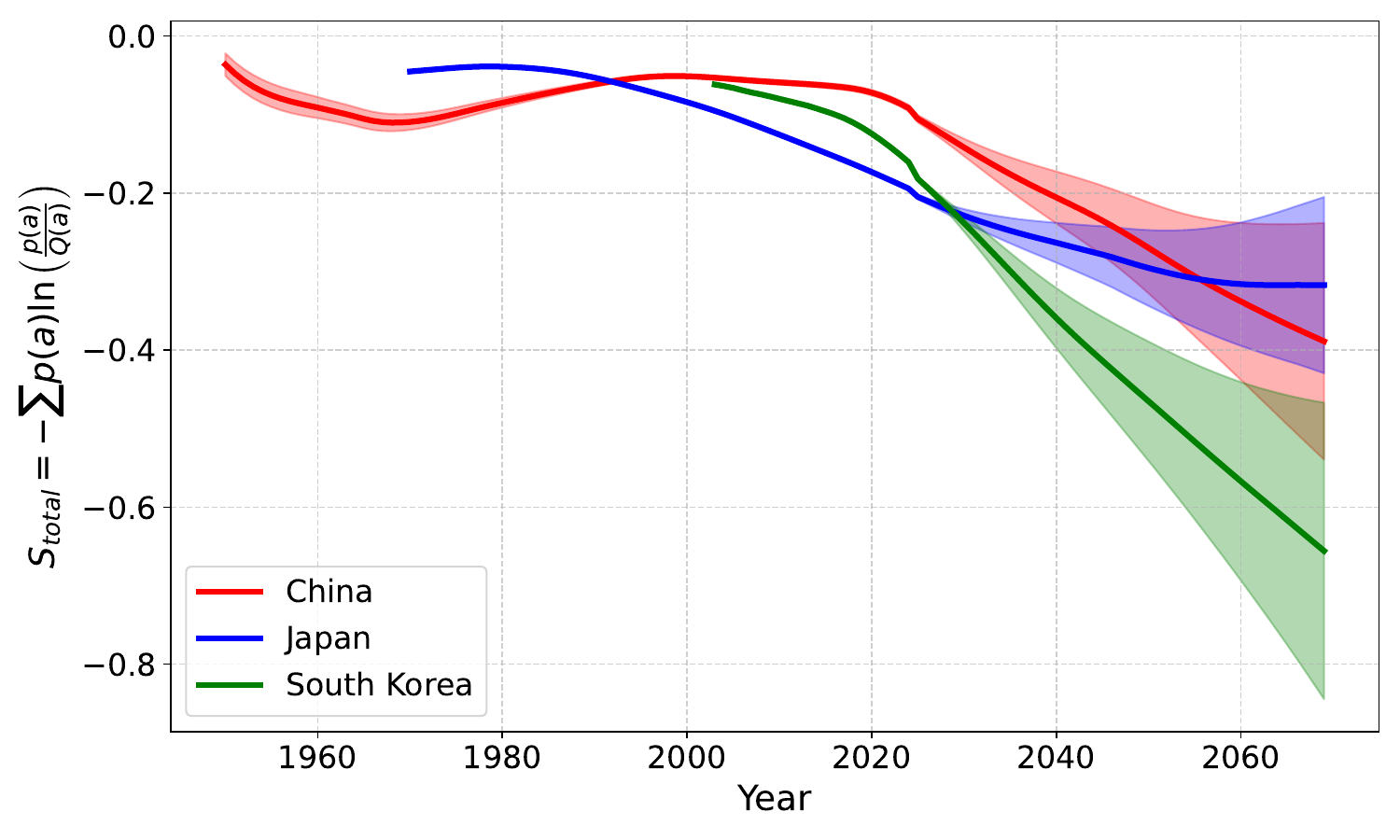}
\caption{Total demographic entropy $S_{\text{total}}(t)$ from 1950 to 2070. The rapid fertility collapse in East Asia drives a steep structural contraction phase, rapidly stripping the populations of their internal potential as they transition into hyper-aged demographic states.}
\label{fig:demographic_entropy}
\end{figure}

\begin{table}[htbp]
\centering
\caption{Non-Equilibrium Physics and Aging Metrics (Mean $\pm$ SD) across Case Studies for Years 2000-2070.}
\label{tab:physics_metrics}
\begin{tabular}{lcccc}
\toprule
Country & Year & Total Entropy $S_{\text{total}}$ & $D_{\text{KL}}$ & Aging Index \\
\midrule
China & 2000 & -0.05(0) & 0.20(2) & 0.27(0) \\
      & 2026 & -0.11(0) & 0.26(5) & 1.11(2) \\
      & 2030 & -0.14(1) & 0.22(6) & 1.59(5) \\
      & 2040 & -0.21(2) & 0.14(7) & 2.95(38) \\
      & 2050 & -0.27(3) & 0.09(8) & 3.43(62) \\
      & 2060 & -0.34(5) & 0.07(7) & 5.42(149) \\
      & 2070 & -0.39(8) & 0.05(7) & 6.08(264) \\
\midrule
Japan & 2000 & -0.08(0) & 0.10(0) & 1.15(0) \\
      & 2026 & -0.21(0) & 0.03(1) & 2.70(2) \\
      & 2030 & -0.23(0) & 0.03(1) & 2.94(7) \\
      & 2040 & -0.26(1) & 0.02(1) & 3.55(31) \\
      & 2050 & -0.29(2) & 0.01(2) & 3.90(55) \\
      & 2060 & -0.32(4) & 0.01(2) & 4.18(83) \\
      & 2070 & -0.32(6) & 0.01(2) & 4.37(123) \\
\midrule
S. Korea & 2026 & -0.19(0) & 0.41(7) & 2.09(1) \\
          & 2030 & -0.24(1) & 0.35(10) & 3.04(9) \\
          & 2040 & -0.36(2) & 0.22(13) & 6.12(103) \\
          & 2050 & -0.47(4) & 0.14(13) & 8.61(255) \\
          & 2060 & -0.57(6) & 0.10(11) & 12.61(532) \\
          & 2070 & -0.67(10) & 0.07(8) & 18.50(1085) \\
\bottomrule
\end{tabular}
\end{table}

As shown in Fig.~\ref{fig:demographic_entropy} and Table~\ref{tab:physics_metrics}, Japan began this structural contraction in the 1980s. China's entropy did not plummet during the strict implementation of the One-Child policy, but in the 2010s, and South Korea exhibits an unprecedented, almost vertical drop starting in the 2000s, reflecting its world-record fertility crash. By 2070, all three systems converge toward the minimal entropy state, structurally locked into rigid, top-heavy distributions with little remaining demographic potential.

Crucially, this entropy contraction acts as a leading indicator of macroscopic demographic decline, preceding both working-age and total population contractions. In Japan, the entropy contraction began in the 1980s, well before the working-age population (15-64) peaked in 1995 (at $\approx 87$~million) and the total population peaked in 2008 (at 128.1~million), which then began its uninterrupted decline in 2010/2011. Similarly, China's entropy contraction accelerated in the 2010s, foreshadowing the peak of its working-age population in 2015 (at just under 1~billion) and its first total population decline in 2022 (peaking at $\approx 1.426$~billion in 2021). For South Korea, the vertical entropy drop of the 2000s anticipated the working-age peak in 2017 (at $\approx 37.6$~million) and the total population decline in 2021 (peaking at $\approx 51.8$~million in 2020). The fact that structural entropy peaks and declines decades before physical population peaks highlights its utility as a critical precursor for identifying systemic demographic turning points.

\subsubsection{Distance from Lotka equilibrium}
In classical mathematical demography~\cite{preston2001}, if the fertility $f(a,t)$ and mortality schedules were frozen at their instantaneous values at year $t$, the population would asymptotically converge to the Lotka stable age distribution:
\begin{equation}
p_t^*(a) = C_t e^{-r_t a} l(a, t),
\end{equation}
where $l(a,t) = \prod_{i=0}^{a-1} s(i, t)$ is the survival probability to age $a$, $r_t$ is the intrinsic rate of natural increase solving the Euler-Lotka equation $\sum_a e^{-r_t a} f(a,t)l(a,t) = 1$, and $C_t$ is the normalization constant. We quantify the non-equilibrium deviation of the actual age structure from this instantaneous stable attractor using the Kullback-Leibler (KL) divergence:
\begin{equation}
D_{\text{KL}}(p(\cdot, t) \parallel p_t^*(a)) = \sum_{a=0}^{A_{\max}} p(a,t) \ln \left( \frac{p(a,t)}{p_t^*(a)} \right).
\end{equation}
This formulation mirrors concepts in non-equilibrium statistical mechanics and information theory, where the KL divergence measures the distance of a transient state from its steady-state Gibbs distribution~\cite{cover2006}, allowing us to treat $D_{\text{KL}}$ as a physical measure of demographic driving force.

For physicists, $D_{\text{KL}}$ quantifies this driving force, indicating how far a rapid change (such as South Korea's birth rate crash) has pushed the system from its steady-state attractor. For policymakers, a high $D_{\text{KL}}$ represents a severe structural mismatch: the current age structure is out of sync with the underlying vital rates. This mismatch implies that even if fertility rates were immediately restored to replacement level, the population would continue to contract for decades due to population momentum (depleted childbearing cohorts). This suggests that immediate policy interventions will have long lags before yielding visible stabilization.

Presently, South Korea shows a massive non-equilibrium deviation ($D_{\text{KL}} \approx 0.449$ in 2024), driven by the rapid collapse of its birth rate. China also exhibits a high divergence ($D_{\text{KL}} \approx 0.290$ in 2024). In contrast, Japan's slow transition keeps it closer to equilibrium ($D_{\text{KL}} \approx 0.034$ in 2024). By 2050, all three countries show a relaxation in $D_{\text{KL}}$ (e.g., dropping to $0.141$ in South Korea), indicating that the bulk age distributions are catching up to their low-fertility attractors.

\subsubsection{Labor scarcity and social support dynamics}
To link these physical state variables to macro-social constraints, we report the Aging Index in Table~\ref{tab:physics_metrics}:
\begin{align}
\text{AgeIdx}(t) &= \frac{\sum_{a=65}^{A_{\max}} \rho(a,t)}{\sum_{a=0}^{14} \rho(a,t)}.
\end{align}
Under the Status Quo forecast, South Korea shows a transition into a hyper-aged, highly ordered state with declining entropy and extremely low potential support ratios. Its Aging Index is projected to reach $860.9\%$ by 2050 (representing over 8.6 elderly citizens for every child), with its PSR dropping to $1.34$. This extreme contraction of the active labor pool suggests that labor scarcity, rather than fiscal resources, becomes the primary physical bottleneck of the social system.

\section{Discussion and conclusions}

In this work, we demonstrate that cohort transport dynamics can be cast as a class of non-equilibrium boundary flux problems where the governing conservation law is exact, but the underlying constitutive relations (fertility and mortality) remain latent and time-varying. By parameterizing these constitutive laws via Bayesian neural networks embedded directly within the transport PDE, our framework guarantees mass conservation while capturing complex, nonlinear age-time interactions. The resulting Bayesian inverse formulation provides a mathematically rigorous way to reconstruct missing state variables from sparse, noisy observations and propagate full posterior predictive uncertainty.

Applying this framework to the CJK triad, we show that these systems undergo a marked structural contraction. By introducing a definition of total demographic entropy, we characterize how age structures collapse into highly ordered, low-entropy states under environmental and socio-economic constraints. The model's validation against independent World Bank data confirms its ability to infer latent vital rates from age-structured transport constraints alone. Specifically, Japan's gradual transition since the 1980s leads to a stable but heavily aged pillar structure. In contrast, South Korea's sharp entropy drop of the 2000s locks the system into a rigid structural state with a projected PSR of 1.34 by 2050. China's transition, accelerating in the 2010s, demonstrates a similar structural contraction, with the working-age cohort peaking in 2015 and total population entering a historical decline in 2022.

Looking forward, the mathematical structure developed here is universal, extending beyond human demographics. Any non-equilibrium system characterized by advective transport, time-varying boundary generation, and internal advective sinks under mass conservation can be modeled using this unified Bayesian inverse framework. Potential applications include tracking the age- or size-structured density of cell divisions and apoptotic sinks in cellular flows and active matter, modeling polymer chain-length degradation and fragmentation under dynamic thermal stresses, and forecasting the structured transport of viral loads within hosts under latent mutation rates and selective pressures. By unifying the flexibility of Bayesian neural networks with the rigorous constraints of physical conservation laws, this framework offers a generalizable paradigm for observing, modeling, and forecasting bounded transport networks across the physical and biological sciences.

\end{document}